\documentclass{svjour2}                   % onecolumn
\smartqed  % flush right qed marks, e.g. at end of proof
\usepackage{graphicx}
\usepackage{mathptmx}      % use Times fonts if available on your TeX system
 \journalname{Experimental Astronomy}
\begin{document}

\title{Development of Ground-testable Phase Fresnel Lenses in Silicon}
% Use \titlerunning{Short Title} for an abbreviated version of
% your contribution title if the original one is too long
\author{John Krizmanic$^{1,2}$, Brian Morgan$^{3}$,
Robert Streitmatter$^{2}$, Neil Gehrels$^{2}$, Keith Gendreau$^{2}$, Zaven Arzoumanian$^{1,2}$, Reza Ghodssi$^{3}$, and Gerry Skinner$^{4}$ }
% Use \authorrunning{Short Title} for an abbreviated version of
% your contribution title if the original one is too long
\institute{J. Krizmanic \at
{NASA Goddard Space Flight Center,} \\
{Greenbelt, Maryland 20771 USA}\\
              Tel.: +001-301-2866817\\
              Fax: +001-301-2861682\\
              \email{jfk@cosmicra.gsfc.nasa.gov}
}
%
% Use the package "url.sty" to avoid
% problems with special characters
% used in your e-mail or web address
%

\date{Received: date / Accepted: date}
% The correct dates will be entered by the editor

\maketitle

\begin{center}
{\it 
$^1$Universities Space Research Association \\
$^2$NASA Goddard Space Flight Center,  Greenbelt, Maryland 20771 USA \\
$^3$Dept. of Electrical and Computer Engineering, University of Maryland, College Park, Maryland 20742 USA \\
$^4$CESR,  9, avenue du Colonel-Roche
31028 Toulouse, FRANCE
}
\end{center}

\begin{abstract}

      Diffractive/refractive optics, such as Phase Fresnel Lenses (PFL's),
  offer the potential to achieve excellent imaging performance
  in the x-ray and gamma-ray photon regimes.  In principle,
  the angular resolution obtained with these devices can be
  diffraction limited.  Furthermore, improvements in signal sensitivity can
  be achieved as virtually the entire flux incident on a lens can be
  concentrated onto a small detector area.  In order to verify experimentally the
  imaging performance, we have fabricated PFL's in silicon using
  gray-scale lithography to produce the required Fresnel profile.
  These devices are to be evaluated in the recently constructed 600-meter
  x-ray interferometry testbed at NASA/GSFC.  Profile measurements of the Fresnel
  structures in fabricated PFL's have been performed and have been used to 
  obtain initial characterization of the expected PFL imaging efficiencies.
 \keywords{Gamma-ray Astronomy \and Optics}
% \PACS{PACS code1 \and PACS code2 \and more}
% \subclass{MSC code1 \and MSC code2 \and more}
\end{abstract}

\section{Introduction}
\label{sec:1}

   The use of Phase Fresnel Lenses (PFL's) offer a mechanism to achieve superb imaging of astrophysical objects in the hard X-ray and gamma-ray energy regimes \cite{Skinner1,Skinner2}.  In principle,
PFL's can concentrate nearly the entire incident photon flux, modulo absorption effects,  with diffraction-limited imaging performance.  The impact of absorption is energy and material dependent, but can be almost negligible at higher photon energies.
The
performance of these diffraction optics is obtained via tailoring the Fresnel profile 
of each zone to yield the appropriate phase at the primary focal point.    However, PFL's have long focal lengths and are chromatic; the excellent imaging is available over a narrow energy range.  In order to demonstrate the imaging capabilities of these optics, we have fabricated ground-testable PFL's in silicon.

%
%
% For figures use
%
\begin{figure}
\centering
% Use the relevant command for your figure-insertion program
% to insert the figure file.
% For example, with the option graphics use
\includegraphics[height=7.5cm]{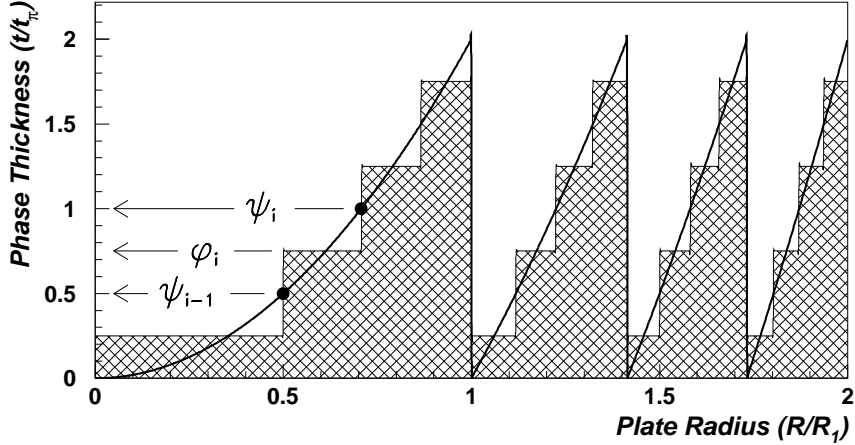}
%
% If not, use
%\picplace{5cm}{2cm} % Give the correct figure height and width in cm
%
\caption{The ideal PFL profile of the first four Fresnel ridges compared to that for a stepped profile with 4 steps/ridge.}
\label{fig:1}       % Give a unique label
\end{figure}

\section{Stepping to PFL's}
\label{sec:2}

   PFL's are a natural extension of Fresnel-diffractive optics.  As opposed to Fresnel Zone Plates (FZP), where alternating half-period or Fresnel zones are completely blocked, and Phase-reversal Zone Plates, where the blocked zone in a FZP is replaced by an amount of material to retard the phase by $\pi$, the entire profile of each Fresnel zone in a PFL has the appropriate amount of material to achieve the proper phase shift at the primary focus.  In practice, the exact profile of a PFL can be approximated by a multiple-step structure as shown in Figure 1 which illustrates the first four Fresnel zones of a PFL with 4 steps approximating the ideal PFL profile.  The location of the radial step transitions is given by
$r_k = r_1 \sqrt{k/P}$ where $r_1 (=\sqrt{2 f \lambda})$ is the location of the first Fresnel ridge ($f$ is the focal length and $\lambda$ the photon wavelength) and $P$ is the number of steps/Fresnel ridge with a step index of $k$.  This choice leads to annuli, as defined by each step, of constant area on the PFL.  Each contributes equally, ignoring absorption, to the intensity at a given focal point.  This configuration of the stepped-phase profile is denoted as a regular-stepped PFL.
   
For an exact PFL profile, all the irradiance appears in the first order ($n=1$) focus.  In a stepped-profile approximation, some energy appears in negative and higher order foci.
Ignoring absorptive effects, the impact on the lens efficiency of approximating the exact PFL profile by a step-wise function is given by \cite{Dammann}
\begin{eqnarray}
Lr_n & =  \biggl{[}\frac{\sin(\pi/P)}{n(\pi/P)}\biggr{]}^2 & ~~{for~ (n-1)/P = m \in Integer} \\
        & = 0  & ~~{otherwise} \nonumber
\end{eqnarray}
where $Lr_n$ is the relative intensity into the $n^{th}$ order focus for $P$ steps in each Fresnel zone.  As $P$ increases, the indices with non-zero intensities of both the real and virtual higher order foci are pushed to higher $n$ with the relative intensities into these higher orders decreasing.  In the limiting case where $P \rightarrow \infty$, the profile is exact for the PFL with the relative intensity in the 1st order ($n=1$) going to 100\% (and the indices of the higher order foci being sent to $\pm \infty$).  More practically, a stepped-PFL with $P=8$ per Fresnel zone has 95\% efficiency focussing into the 1st order focal point, {\it sans} absorption.

  The material needed in a PFL to retard the phase also will attenuate the flux of incident photons. The index of refraction of matter can be expressed as $n^\ast = 1 - \delta -i\beta$ and is related to atomic scattering factors \cite{Henke}.  Thus for a photon of wavelength $\lambda$, a material of thickness $t$ will retard the phase by $\varphi = 2\pi t \delta/\lambda$ while attenuating the intensity by $e^{-t/\tau}$ where $\tau = (4\pi\beta)^{-1}$.  The attenuating effects of the material in a Fresnel zone of a stepped-PFL can be calculated by determining the amplitude of the waveform traversing each step of the PFL profile.  If $t_i$ is the material thickness of the $i^{th}$ step in a particular Fresnel zone, the phase will be retarded by $\varphi_i$.  As shown in Figure 1, the $i^{th}$ step retards the phase between  $\psi_{i-1}$ and  $\psi_i$, and the amplitude can be expressed as \cite{Kirz}
\begin{eqnarray}
A_i & = & \frac{C}{2\pi} e^{-t_i/2\tau} \int_{\psi_{i-1}}^{\psi_i} e^{i(\psi-\varphi_i)}d\psi \\
      & = & \frac{C}{2\pi} e^{-t_i/2\tau}(\alpha^R_i - i\alpha^I_i)
\end{eqnarray}
where $C$ is a normalization constant and 
\begin{eqnarray*}
\alpha^R_i & = & e^{-t_i/2\tau} \cos{\varphi_i}(\sin{\psi_i}-\sin{\psi_{i-1}}) - \sin{\varphi_i}(\cos{\psi_i}-\cos{\psi_{i-1}}) \\
\alpha^I_i & = & e^{-t_i/2\tau} \cos{\varphi_i}(\cos{\psi_i}-\cos{\psi_{i-1}}) + \sin{\varphi_i}(\sin{\psi_i}-\sin{\psi_{i-1}})
\end{eqnarray*}
Summing over all $P$ steps leads to determining the intensity at the primary focus
\begin{eqnarray}
I_1 &=& |A|^2 = AA^\ast = \frac{C^2}{4\pi^2}  \biggl(\biggl[\sum_{i=1}^P \alpha_i^R \biggr]^2 + \biggl[\sum_{i=1}^P\alpha_i^I \biggr]^2\biggr)
\end{eqnarray}
Note that circular symmetry is assumed for the PFL, and this calculation is for a single Fresnel zone.  If a PFL contains a total of $M$ individual Fresnel zones with identical, in phase, profiles, the irradiance at the focus would be increased by $M^2$ as each Fresnel zone has the same area on the PFL.
This formulation holds for any step spacing, regular or irregular, as long as a sufficiently small scale exists where the phase thickness is effectively constant.  Choosing an energy of 8 keV, the efficiency of a $P=8$ regular-stepped PFL, including absorption, is 82.3\% in silicon. If absorption is ignored, the efficiency is 95\% which is exactly that as determined from Equation 1 for $n=1$.

\section{Ground-test Constraints and PFL Fabrication}

For a PFL with diameter $d$, minimum Fresnel ridge spacing $p_{min}$, focusing at a photon wavelength $\lambda$, the focal length is given by
\begin{eqnarray}
f & = & \frac{p_{min}d}{2\lambda} \\
   & \approx & 4 ~\frac{p_{min}}{\mu {\rm m}}~\frac{d}{\rm cm}~\frac{E}{\rm keV} 
\end{eqnarray}
where $f$ is in meters for $p_{min}$ in $\mu$m, $d$ in cm, and $E$ in keV in Equation 6.  Using the representative values of $p_{min} = 25 ~\mu$m, $d = 1$ cm, and $E = 8$ keV (Cu K-$\alpha$), the focal length would be 800 meters which is rather long for a ground-test.

At NASA Goddard Space Flight Center, a 600-meter Interferometry Testbed is available for testing of PFL optics.  The nominal configuration has an optics station 150 meters from an x-ray source and a detector station 450 meters from the optics station.  Assuming the x-ray emission is isotropic within the field-of-view of the optics, the effective focal length of an optic focussing at the detector station would be $f_{Eff} = \frac{f_1 f_2}{f_1+f_2} =112.5$ meters.  This sets the value of the focal length of a PFL for incorporation into this test beam configuration.  Using $f=112.5$ meters, $d=1$ cm, and $E = 8$ keV, this leads to a minimum Fresnel ridge spacing of $3.5~\mu$m which is the natural scale size for micro-fabrication in silicon.  The Fresnel ridge height needed to retard the phase by $2 \pi$ is given by $t_{2\pi} = \lambda/\delta$ where $\lambda$ is the photon wavelength and $\delta$ is the real part of the index of refraction.  For silicon, $t_{2\pi} \approx 2.57 ~\bigl[E_\gamma/{\rm keV}\bigr]$ $\mu$m or 20.5 $\mu$m at 8 KeV.

\subsection{PFL fabrication using Gray-scale lithography}

 The gray-scale lithographic fabrication process has been employed at the University of Maryland to create PFL structures in silicon wafers \cite{Morgan1}.  This implementation of the gray-scale process employs a lithographic mask that uses small, variable-transmission pixels (gray levels) that create, via projection photolithography, a designed, 3-dimensional structure in a photoresist spun on a silicon wafer.  This pattern is then transferred into the silicon via deep-reactive ion etching (DRIE).  The developed ground-test PFL's have been fabricated using silicon-on-insulator (SOI) wafers in order to minimize the thickness of the required mechanical substrate under the PFL's and thus maximize the x-ray photon transmission.  The sandwiched oxide layer forms a natural etch stop to remove the silicon substrate directly under each PFL while leaving the surrounding material for mechanical stability.  The unprocessed SOI wafer was 100 mm in diameter with 70 $\mu$m of silicon, 2 $\mu$m oxide, and 500 $\mu$m silicon forming the SOI wafer structure.  A prototype silicon PFL fabricated using the gray-scale lithographic process is shown in Figure 2.
 
%
%
% For figures use
%
\begin{figure}[t]
\centering
% Use the relevant command for your figure-insertion program
% to insert the figure file.
% For example, with the option graphics use
\includegraphics[height=6cm]{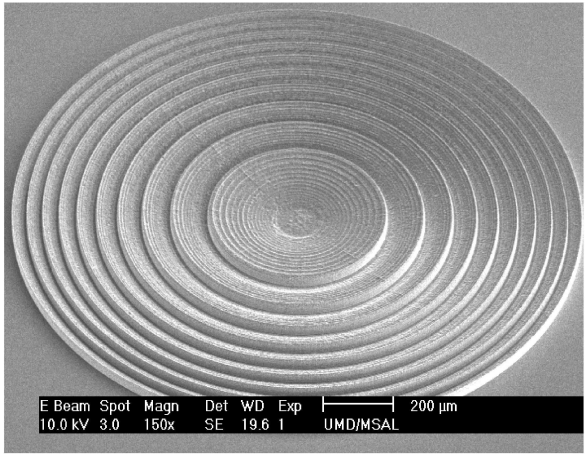}
%
% If not, use
%\picplace{5cm}{2cm} % Give the correct figure height and width in cm
%
\caption{A prototype PFL fabricated using the gray-scale lithographic process[6].
Copyright 2004 IEEE.}
\label{fig:2}       % Give a unique label
\end{figure}

%
% For tables use
%
\begin{table}
\centering
\caption{Ground-test PFL design parameters.  These devices are designed to have a Fresnel ridge height of approximately 40 $\mu$m corresponding to a $4 \pi$ phase shift.}
\label{tab:1}       % Give a unique label
%
% For LaTeX tables use
%
\begin{tabular}{ccccc}
\hline\noalign{\smallskip}
PFL Designation & Diameter & $p_{min}$ & \# of Ridges & \# Steps/Ridge  \\
\noalign{\smallskip}\hline\noalign{\smallskip}
X3 & 2.99 $mm$ & 24 $\mu m$ & 32 & 16 \\
X4 & 2.99 $mm$ & 24 $\mu m$ & 32 & 16 \\
X5 & 2.99 $mm$ & 24 $\mu m$ & 32 & 8 \\
X6 & 4.72 $mm$ & 15 $\mu m$ & 80 & 8 \\
\noalign{\smallskip}\hline
\end{tabular}
\end{table}
 
 Table 1 lists the four PFL designs that have been included in this ground-test fabrication.  Note that this PFL fabrication incorporated a design to produce $4 \pi$, as opposed to $2 \pi$ thick Fresnel optics.  This was chosen for this initial fabrication to effectively double the minimum ridge spacing, p$_{min}$, for a set focal length, PFL diameter, and design energy.  Although this will increase absorption losses, the relaxation of the $p_{min}$ requirement eased the constraints on the device fabrication.  The four PFL's, along with several test structures, were grouped to form a {\it die} which is compact in spatial extent.  Twelve of these {\it dice} in a $3 \times 4$ array were fabricated on the 100 mm SOI wafer via a step-and-repeat process. 
 
 The goal of this fabrication was to produce a sample of PFL's for testing in the 600 m beam line, and the process was not optimized for yield.  In order to identify the optimal PFL's for testing, an optical inspection rejected those with obvious defects. This rejected 15 out of the possible 48 PFL's.  The remaining PFL's were scanned via an optical profilometer (Veeco, WYKO NT1100) to determine the accuracy  of the fabricated profiles.  For the 3 mm diameter PFL's, the first and last 5 Fresnel ridges were scanned and compared to the design profile.  For the 5 mm PFL, the 5 ridges near the half radius were also scanned and compared.  Using an analysis similar to that presented in Equation 4, albeit ignoring absorption and using a phasor formalism, the efficiency of each scanned PFL was estimated from the profiles obtained from the profilometer measurements.  Note that the profiles are measured along a chosen radial path and circular symmetry was assumed.  Figure 3 illustrates the profile measurements and a comparison to the design profile for a 3 mm diameter PFL (X3) for the regions near the center of the device (leftmost plot) and near the edge (rightmost plot).

\section{PFL Anticipated Performance}
 
%
%
% For figures use
%
\begin{figure}
\centering
% Use the relevant command for your figure-insertion program
% to insert the figure file.
% For example, with the option graphics use
\includegraphics[height=6cm]{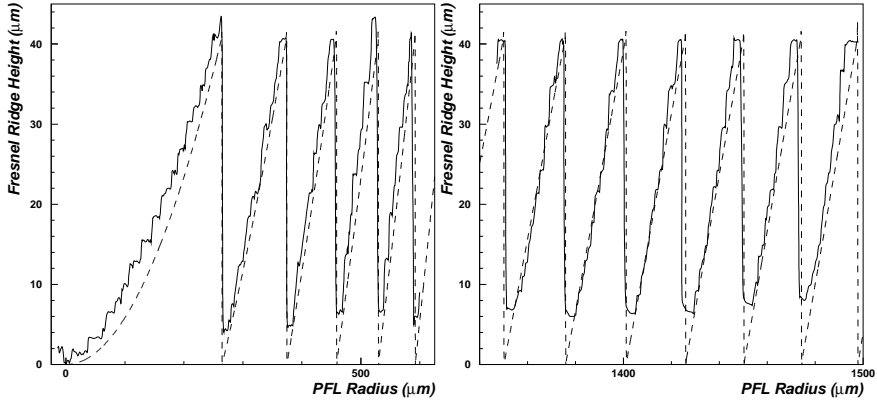}
%
% If not, use
%\picplace{5cm}{2cm} % Give the correct figure height and width in cm
%
\caption{The measured profile (solid) of a fabricated, 3 mm diameter (X3) PFL as compared to the design profile (dashed).  The left plot shows the results at the center of the lens while the right plot illustrates the results near the outermost region of the lens.  These measurements lead to anticipated efficiencies of 76\% for the center region and 64\% near the edge of the PFL.}
\label{fig:3}       % Give a unique label
\end{figure}
%
% For tables use
%
\begin{table}
\centering
\caption{Fabricated PFL anticipated efficiencies obtained from profile measurements and based upon PFL device type.  The maximum and minimum designations are the efficiencies determined for a specific, fabricated PFL.}
% as opposed to the profile measurement location.}
\label{tab:1}       % Give a unique label
%
% For LaTeX tables use
%
\begin{tabular}{ccccccc}
\hline\noalign{\smallskip}
 & \multicolumn{2}{c}{X3} & \multicolumn{2}{c}{X4} & \multicolumn{2}{c}{X5} \\
 & Center & $r=1.5~mm$ & Center & $r=1.5~mm$ & Center & $r=1.5~mm$ \\
\noalign{\smallskip}\hline\noalign{\smallskip}
Average & 75.6\% & 59.0\% & 72.5\% & 54.8\% & 68.5\% & 55.3\% \\
Maximum &77.2\% & 67.5\%& 80.9\% &64.8\% &74.8\% & 63.4\%  \\
Minimum & 69.6\%& 48.0\% & 52.6\% & 41.1\% & 59.1\% & 36.5\%   \\
\noalign{\smallskip}\hline
\end{tabular}
\begin{tabular}{cccc}
\hline\noalign{\smallskip}
 & \multicolumn{3}{c}{X6} \\
 & Center & $r=1.5~mm$ & $r=2.5~mm$  \\
\noalign{\smallskip}\hline\noalign{\smallskip}
Average & 61.6\% & 55.1\% & 32.4\% \\
Maximum & 65.2\% & 61.6\% & 36.1\% \\
Minimum & 35.8\% & 54.8\%& 21.1\% \\
\noalign{\smallskip}\hline
\end{tabular}
\end{table} 

 Table 2 lists the maximum, minimum, and average efficiency for the different fabricated PFL's based upon the profile measurements.  The values for the maxima and minima quoted for a PFL are that for a specific lens,  i.e. the ensemble of measurements for a specific design were used to determine the appropriate designation. The quoted efficiencies do not take into account absorptive losses due to either the Fresnel profile or the $\sim25~\mu$m substrate.  Assuming an 8 step/Fresnel ridge profile and 8 keV, the reduction in collection efficiency is approximately 14\%, i.e. $1-\frac{0.82}{0.95}$, due to the phase-retarding material in the stepped-Fresnel profile and 30\%, i.e.  $1-e^{-25~{\mu m}/70~{\mu m}}$, due to the $25~\mu$m silicon substrate.  Note that the effects of attenuation can be significantly reduced by fabricating PFL's designed for higher photon energies.
 
 The data represented in Table 2 demonstrate that, as indicated from profile measurements, stepped-profile PFL's micro-fabricated in silicon have efficiencies significantly larger than that for the simpler zone plates and phase-reversal zone plates.  The data also illustrate that efficiencies determined from the finer pitch Fresnel zones are reduced as compared to the larger pitch center Fresnel zones.  This is due to the fact that it is more difficult to accurately fabricate zones with higher aspect ratios, defined as the ratio of Fresnel ridge height to ridge pitch.  A significant contribution to this effect is due to the aspect-ratio dependence of the etching process; it is more difficult to remove silicon from narrow ridge regions as shown in Figure 3.  Work has progressed on designing appropriate compensation in the lithographic mask and this technique has been demonstrated in the fabrication of a second-generation of Fresnel Lens structures that exhibit a much reduced aspect-ratio dependence of the PFL profiles \cite{Morgan2}.
 
  There are three components contributing to the angular resolution of a PFL: diffraction, detector spatial resolution, and chromatic aberration.  For the 3 mm ground-test PFL imaging at 8 keV, the  values of each of these terms is given by
\begin{eqnarray}
{\rm Diffraction~Limit:}~~~ & \vartheta_D = & 1.22 \lambda/d = 8 ~{\rm milli\!-\!arcseconds}~(m^{\prime \prime}) \nonumber  \\
{\rm Detector ~Spatial ~Limit:} ~~~& \vartheta_s = & \Delta x/F = 6 ~m^{\prime \prime}  \\
{\rm Chromatic ~Aberration ~Limit{[1]}:}~~~& \vartheta_{\Delta E} = & 0.2 (\Delta E/E)(d/F) = 5 ~m^{\prime \prime} \nonumber 
\end{eqnarray}
where a $\Delta x =13$ $\mu$m detector pixel size is assumed with $\Delta E = 140$ eV FWHM and $F=450$ meters.  Note that the contribution to the angular resolution from the chromatic aberration term is reduced if one assumes the Cu K-$\alpha$ line width.  Thus the anticipated angular resolution of these ground-test PFL's are $\sim 10$ milli-arcseconds ($m^{\prime \prime}$) which is a significant improvement to that obtained from current astronomical missions, e.g. $500~m^{\prime \prime}$ for CHANDRA \cite{Chandra}, in this energy range.

\section{Conclusions}

 We have fabricated ground-test PFL's in silicon using gray-scale lithography.  We have determined the imaging performance of these devices via analysis of the measured profiles of the fabricated optics.  These results indicate that the efficiencies, although less than ideal, are a significant improvement over the theoretical maximum that can be obtained with zone plates and phase-reversal zone plates.
We plan on introducing these devices into the 600 m test beam to demonstrate their imaging capability and verify the anticipated efficiency determination via {\it in situ} x-ray measurements.
This material is based upon work supported by the National Aeronautics and Space Administration under Grant APRA04-0000-0087 issued through the Science Mission Directorate Office and by Goddard Space Flight Center through the Director's Discretionary Fund.
%
%
% BibTeX users please use
% \bibliographystyle{}
% \bibliography{}
%
% Non-BibTeX users please follow the syntax
% the syntax of "referenc.tex" for your own citations
%\input{referenc}
%%%%%%%%%%%%%%%%%%%%%%%%%%%%%%%%%%%%%%%%%%%%%%%%%%%%%%%%%%%%%%%%%%%%%%  }

% Non-BibTeX users please use

%%%%%%%%%%%%%%%%%%%%%%%%%%%%%%%%%%%%%%%%%%%%%%%%%%%%%%%%%%%%%%%%%%%%%%

\end{document}